\documentclass[10pt, conference]{IEEEtran}

\usepackage{amssymb}
\usepackage{amsmath}

\usepackage[colorlinks=true, linkcolor=blue, citecolor=blue, urlcolor=blue]{hyperref}

\usepackage{graphicx}
\usepackage{booktabs}
\usepackage{multirow}
\usepackage{tabularx}
\usepackage{enumitem}
\usepackage{threeparttable}
\usepackage{balance}
\usepackage{url}

\usepackage{fancyhdr} 
\fancypagestyle{firststyle}
{
\fancyhf{}
\fancyfoot[C]{\scriptsize{Proceedings of the IEEE International Conference on Software Analysis, Evolution and Reengineering (SANER 2025), Montreal, IEEE, pp.~891--896. \\ This version is the author's copy. The publisher's definite version is available online via \url{https://doi.org/10.1109/SANER64311.2025.00097}.}}
}

\begin{document}

\title{An Empirical Study of Vulnerability \\ Handling Times in CPython}

\author{
\IEEEauthorblockN{Jukka Ruohonen}
\IEEEauthorblockA{University of Southern Denmark \\
Email: juk@mmmi.sdu.dk}
}

\maketitle

\begin{abstract}
The paper examines the handling times of software vulnerabilities in CPython,
the reference implementation and interpreter for the today's likely most popular
programming language, Python. The background comes from the so-called
vulnerability life cycle analysis, the literature on bug fixing times, and the
recent research on security of Python software. Based on regression analysis,
the associated vulnerability fixing times can be explained very well merely by
knowing who have reported the vulnerabilities. Severity, proof-of-concept code,
commits made to a version control system, comments posted on a bug tracker, and
references to other sources do not explain the vulnerability fixing times. With
these results, the paper contributes to the recent effort to better understand
security of the Python ecosystem.
\end{abstract}

\begin{IEEEkeywords}
Python, defect, software security, software vulnerability, vulnerability
disclosure, bug fixing times, bug severity
\end{IEEEkeywords}

\section{Introduction}

\thispagestyle{firststyle} 

Underneath the Python programming language are so-called virtual machines that
compile Python code into byte code before execution. These machines are embedded
into the language's interpreters. While there are many interpreters, including
examples such as Cython and Jython, the reference implementation, CPython, is
the most popular one. Over the years, many vulnerabilities have also been
reported for this reference implementation~\cite{LinHua22}. The present paper
examines how long the handling of these have taken, what factors explain the
handling times, and how well these can be predicted.

By handling it is meant that addressing of vulnerabilities requires many
distinct software engineering work tasks. A~vulnerability needs to be obviously
fixed, but a given fix needs to be also integrated into releases, often
including distinct release branches. In addition, the vulnerability requires
coordination between multiple parties~\cite{Lin23, Ruohonen18IST}, which in the
open source context include particularly so-called downstream distributors, such
as Linux distributions. Coordination is also required with the non-profit MITRE
corporation to get a Common Vulnerabilities and Exposures (CVEs)
identifier. Although the Python Software Foundation (PSF), who as an
organization is behind CPython, is a CVE numbering authority and can thus
allocate CVEs on its own, it may be that additional coordination is still
required with some CVEs before they are published by MITRE and later on archived
into the National Vulnerability Database (NVD)~\cite{NVD24}, the world's
foremost vulnerability database. For these reasons, the paper concentrates on
two distinct timelines within a vulnerability's overall handling time: (a)~the
time required to fix a given vulnerability and (b)~the time required for a CVE
for it to be published. Hereafter, the former is known as \textit{fixing time}
and the latter as \textit{CVE coordination time}.

The questions examined and the paper's topic in general are easy to
motivate. According to benchmarks, Python is the most popular programming
language today~\cite{Cass24}, and because CPython is the most popular
interpreter for the language, the vulnerabilities affecting the interpreter
affect large user and deployment bases. In addition, as pointed out in the
opening Section~\ref{sec: related work}, the handling times proxy not only
software engineering effort but also security risks. A further motivating point
is that the paper's topic has not been examined previously, despite a large
reference literature base on bug and vulnerability handling times, including
their fixing times. In this regard, the paper contributes to the existing and
related work by patching a small but notable knowledge gap.

The paper's remaining structure is simple. After the already noted
Section~\ref{sec: related work} on related work, the dataset and methods for
examining it are elaborated in Section~\ref{sec: data and methods}. Then, the
empirical results are presented in Section~\ref{sec: results}. Finally,
Section~\ref{sec: discussion} summarizes the conclusions reached, pinpoints some
limitations, and discusses the implications particularly for further work.

\section{Related Work}\label{sec: related work}

There are two rather large branches of related work. The first branch is
sometimes known as a vulnerability life cycle analysis~\cite{Frei06,
  Marconato13}. Like normal, non-security bugs, vulnerabilities are discovered,
reported, coordinated, fixed, and archived to databases, among other
things. These and other events that occur during a vulnerability's life cycle
allow to formulate different longitudinal research questions and setups for
these.

For instance, a time difference between a date of discovering a vulnerability
and a date when a discoverer first contacted a vendor affected by the
vulnerability allows to approximate communication delays and potential
communication obstacles in vulnerability disclosure~\cite{Ruohonen20CHB}. As
elaborated in Section~\ref{subsec: variables}, the contacting dates, often also
known as vulnerability disclosure dates, are important also in the present work
because they set operational reference points for both timelines considered.

The PSF has also a specific vulnerability disclosure policy. In essence, a
discoverer should privately contact a security team at the PSF, who then
determines whether the issue reported is really a vulnerability, and if so,
handles the required coordination with the discoverer privately, contemplates
whether the vulnerability can be publicly discussed in a bug tracker, and then
fixes the vulnerability, integrates the fix developed to releases, and releases
security advisories for the vulnerability~\cite{PSF24a}. Against this backdrop,
the fixing times considered proxy particularly the software engineering work
required. Although not perfectly, both the fixing times and the CVE coordination
times proxy also security risks; a long delay imply more potential for
exploitation of a given vulnerability.

Similar timelines have been widely used in previous work~\cite{Ruohonen20CHB,
  Ruohonen18IST, Ruohonen19RSDA}. The patch development aspect and the software
engineering tenet imply that the branch and the paper too further interlace with
a closely related empirical research domain that has examined and predicted bug
fixing times~\text{\cite{Abdelmoez12, Akbarinasaji18, Schroter10,
    ZhangKhomh12}}. While also this domain essentially operates with time
differences, bug tracking systems have prompted also more convoluted questions
between a bug's state changes, including a question of which bugs get reopened
in bug tracking systems~\cite{Zimmermann12}. In contrast, the vulnerability life
cycle branch usually, either explicitly or implicitly, maintains that a
vulnerability's life cycle is more or less a linear process.

The second research branch originates from the Python programming language
itself. In particular, a lot of work has been done in recent years to examine
the security of software written in Python. In addition to vulnerability
detection in Python code~\cite{WangXu24}, the branch has operated with an
ecosystem-wide scope, examining particularly the packages distributed in the
Python's PyPI repository. Here, on one hand, many Python packages have been
observed to suffer from various software quality issues, including real and
potential security flaws~\text{\cite{Alfadel23, Ruohonen21PST}}. On the other
hand, it has also been observed and argued that a probability of picking a safe,
non-vulnerable package from PyPI is still relatively high~\cite{Paramitha23}. To
some extent, this observation aligns with results from time series analysis;
only a recent past has been observed to be relevant for predicting a probability
that a Python package's current version is
vulnerable~\cite{Ruohonen18IWESEP}. By and large, but not
entirely~\cite{LinHua22}, this largely empirical branch of research has
overlooked a fact that over the years a lot of vulnerabilities have been
reported also for the language's core, CPython itself. Because this reference
implementation is written in the classical C language, also the vulnerabilities
are somewhat or even very different than vulnerabilities in packages written in
Python.

\section{Data and Methods}\label{sec: data and methods}

\subsection{Data}

Although there are many tools for extracting data from software repositories,
including those written in Python~\cite{Spadini18}, the relatively small amount
of vulnerabilities for CPython allowed to construct the dataset
manually. Another, more practical reason for the manual construction is that
CPython recently moved from a custom tracker \cite{PSF24b} to a more systematic
one using the Open Source Vulnerability (OSV) format~\cite{PSF24c}. The latter
is hosted on GitHub. For the purposes of this paper, the old tracker had a
benefit in that it recorded many distinct vulnerability handling dates
explicitly. For this reason, the data contains only the $n = 93$ vulnerabilities
that were fully present in the old tracker. A few vulnerabilities from 2023 had
to be excluded from the old tracker because they were not fully recorded due to
the associated migration to the new tracker. Finally, it should be stressed
that not all of the vulnerabilities that have affected CPython have been about
the interpreter \textit{per~se}. As CPython bundles other software components,
their vulnerabilities affect also the interpreter and its security. The Expat
library is the prime example in this regard.

\subsection{Variables}\label{subsec: variables}

\subsubsection{Dependent Variables}

Two separate dependent variables are used for the empirical analysis. The first
is:
\begin{equation}\label{eq: patching}
(\textit{Fixing Time})_i
= (\textit{Last Commit})_{it_a} - (\textit{Disclosure})_{it_c} ,
\end{equation}
where $i = 1, \ldots, n$ refers to the $i$:th vulnerability, $(\textit{Last
  Commit})_{it_a}$ to the last commit in a version control system that was made
to fix the $i$:th vulnerability at date $t_a$, regardless of a branch, and
$(\textit{Disclosure})_{it_c}$ to a date $t_c$ at which the $i$:th vulnerability
was first disclosed to the PSF's security team or otherwise made known to the
CPython's developers according to the meta-data from the old vulnerability
tracker. If multiple persons reported a same vulnerability, the officially
recorded disclosure date is still used. It should be also remarked that the
CPython's developers may have identified and categorized multiple
vulnerabilities identified with a single CVE. In any case, $(\textit{Fixing
  time})_i$ is a typical count data variable approximating particularly the
software engineering effort required to fix vulnerabilities.

The second dependent variable is operationalized as:
\begin{align}
(\textit{CVE Coordination Time})_i
&= (\textit{CVE Publication})_{it_b} \\
\notag &- (\textit{Disclosure})_{it_c} ,
\end{align}
where $(\textit{CVE Publication})_{it_b}$ is a date $t_b$ at which a CVE was
published for the $i$:th vulnerability according to the meta-data from the old
tracker. Four remarks are necessary about this variable. First, the
vulnerabilities, as identified as such by the CPython's developers, lacking CVEs
had to be obviously excluded. Second, in case two or more CVEs were allocated
for a single vulnerability, as identified as a single vulnerability in the old
tracker, the one with the earliest date was used. Third, a restriction
$(\textit{CVE Coordination Time})_i \geq 0~\forall~i$ was imposed, meaning that
those vulnerabilities were excluded that had CVEs allocated already before the
associated disclosure dates. These cases have supposedly happened due to a given
discoverer or other reporter having obtained a CVE from MITRE on his or her own,
without first contacting the PST's security team. Fourth, as the variable's name
indicates, it is taken to proxy coordination of CVE identifiers, but it is also
important in terms of security because many companies allegedly only deploy
security patches fixing vulnerabilities identified with CVEs. Much work has also
been done to help companies and others with patching
prioritization~\text{\text{\cite{Kocaman22, Yadav22}}}.







\subsubsection{Independent Variables}

the independent, explanatory variables are as follows:
\begin{enumerate}
\itemsep 5pt
\item{REPORTER: a set of dummy variables identifying reporters of the CPython
  vulnerabilities. Note that a reporter may or may not be the same person who
  originally discovered a given vulnerability. Reporters were identified by
  their unique real names or pseudonyms. The rationale for this independent
  variable builds on many existing studies, which have hypothesized and observed
  that reporters' and discoverers' skills and characteristics, including their
  communication skills, affect bug and vulnerability handling
  times~\text{\cite{Akbarinasaji18, Giger10, Marks11, Ruohonen20CHB,
      ZhangKhomh12}}.}
\item{SEVERITY: severity of the vulnerabilities patched in CPython, as measured
  with a $[0, 10]$ interval-scaled variable based on the the Common
  Vulnerability Scoring System (CVSS~v.~3) information; the higher a value, the
  more severe a vulnerability. If a CVE was missing, a value zero was used, and
  if NVD lacked CVSS data for a CVE, the CVSS (v.~2) information from the old
  tracker was used. Although existing results have not often been
  confirmatory~\text{\cite{Alfadel23, Marks11, Ruohonen20CHB}}, the severity of
  bugs and vulnerabilities has been a classical variable used to model and
  predict the associated handling times~\text{\cite{Abdelmoez12, Akbarinasaji18,
      Giger10, ZhangKhomh12}}. However, the hypothesized direction of an effect
  is not entirely clear. On one hand, as developers may prioritize severe
  vulnerabilities, their patching times should be shorter. On the other hand,
  particularly severe vulnerabilities may be time-consuming to analyze and
  fix. In any case, in practice the CVSS (v.~3) severity information was
  obtained from hyperlinks in the CPython's old tracker pointing directly to the
  NVD.}
\item{POC: a dummy variable indicating whether a reporter or some other person
  had included a proof-of-concept (POC) code for demonstrating a given
  vulnerability. The recording only counts POCs that were explicitly, as visible
  code, posted to the initial bug report for a vulnerability. In other words,
  POCs referenced by hyperlinks to external sources were excluded. The
  variable's rationale comes from existing research on bug fixing times; POCs,
  reproducible tests, stack traces, screenshots, and associated things have been
  observed to shorten bug (and vulnerability) fixing times~\cite{Karim19,
    Ruohonen20CHB, Schroter10}. The reason seems clear: the more there is robust
  information available, the easier and hence faster it is to fix a
  vulnerability or a non-security bug.}
\item{COMMITS: the number of commits to a version control system that were
  required to patch a given vulnerability in all branches. The rationale is
  straightforward: a large amount of commits and thus effort should lengthen the
  timelines. If many commits were required, a vulnerability may have been
  particularly complex, difficult to interpret and debug, or otherwise
  time-consuming to fix. Although existing results are not entirely
  definite~\cite{Ruohonen19RSDA}, code churn has also been observed to lengthen
  bug fixing times~\cite{ZhangKhomh12}.}
\item{REFERENCES: a number of online references to external sources, counted in
  terms of hyperlinks posted to a given bug report. Although internal hyperlinks
  pointing to a tracker's other bug reports or other elements were excluded
  together duplicate hyperlinks, the variable counts also hyperlinks posted by
  bots, including those referencing commits. The manually posted hyperlinks
  typically point toward bug trackers and other tracking systems used by other
  projects, including operating system vendors, although references to Python
  online documentation, standards, mailing lists, blogs, security companies, and
  many other things are present too. A similar variable has been used also in
  previous studies~\cite{Ruohonen18IST}. The rationale is that the vulnerability
  handling timelines might be shorter for vulnerabilities that are widely
  discussed and popularized on different communication channels and
  platforms. Though, it may also be that a widely discussed vulnerability is
  controversial or otherwise problematic, perhaps indicating a longer handling
  time already due to the time required to discuss and collectively interpret
  it.}
\item{COMMENTS: by again following existing research~\text{\cite{Giger10,
      ZhangKhomh12}}, the number of comments that were posted to an initial bug
  report for a vulnerability. The rationale is similar to REFERENCES. It can be
  additionally remarked that the actual content of discussions on bug reports
  does not seem to affect bug fixing times~\cite{Noyori19}. Therefore, merely
  including a variable measuring a volume of discussions seems justified.}
\end{enumerate}

Four remarks are in order about these independent variables. First, it should be
mentioned that during the period observed CPython had also used two bug tracking
system, an internal one and a one hosted on GitHub. The presence of two trackers
complicates the operationalization of COMMENTS and REFERENCES. In terms of the
former variable, all comments were accounted for vulnerabilities that were
handled in the internal tracker, whereas in the GitHub case the
operationalization only counted comments identified as such by GitHub, meaning
that mentions, references, and other entries in GitHub's parlance were
excluded. Then, in terms of REFERENCES, in addition to the exclusion of internal
references, also cross-references between the two trackers were
excluded. Second, in case a bug report was not referenced in the old tracker,
POC, REFERENCES, and COMMENTS were all scored with a value zero. Third, even
though hypothesized effects are not entirely unambiguous, all variables are
well-justified in terms of existing research. Fourth and last, only six
variables were operationalized even though plenty more would be easy to derive
and operationalize. The reason is the small sample~size.

\subsection{Methods}

Broadly speaking, the literature has operated with two methodological
approaches: classification and regression analysis. The former approach
typically splits bug and vulnerability handling times into two categories;
``short'' and ``long'' or ``fast'' and ``slow'' ~\cite{Abdelmoez12,
  Akbarinasaji18, Giger10, Marks11, ZhangKhomh12}. This approach suffers from an
obvious limitation that a threshold for a split is more or less arbitrary. The
second approach treats a patching time as a continuous count data
variable. Therefore, the typical methodological choices include ordinary least
squares with variable transformations~\cite{Ruohonen20CHB}, Poisson regression,
negative binomial regression~\cite{Ruohonen21PST}, quantile
regression~\cite{Ruohonen18IST}, and Cox's proportional hazards
regression~\cite{Ruohonen19RSDA}. The regression approach is used in the present
work. Two estimators are used: a standard ordinary least squares (OLS) regression
with a $\ln(x + 1)$ transformation for the two dependent variables and a
so-called Huber's $M$-estimator with the same transformation. The latter belongs
to a family of robust regression methods, and, therefore, it is generally much
less sensitive to outliers. Without delving into the statistical details, which are well-documented~\cite{deMenezes21}, the $M$-estimator is computed with the \texttt{rlm} function from the \texttt{MASS} library for the R language.

\section{Results}\label{sec: results}

\begin{figure}[th!b]
\centering
\includegraphics[width=\linewidth, height=3cm]{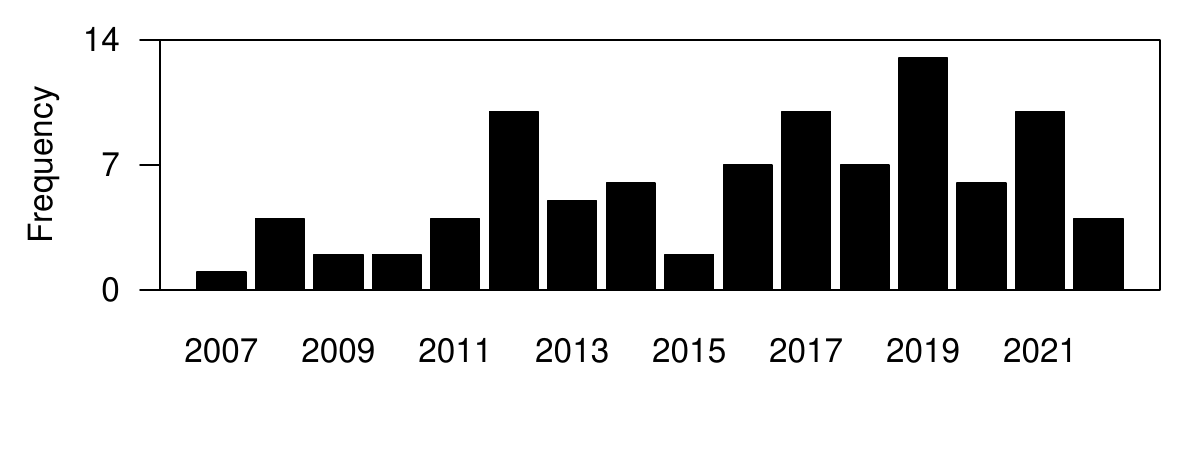}
\caption{Publication Years According to Disclosure Dates}
\label{fig: years}
\end{figure}

\begin{figure}[th!b]
\centering
\includegraphics[width=\linewidth, height=3cm]{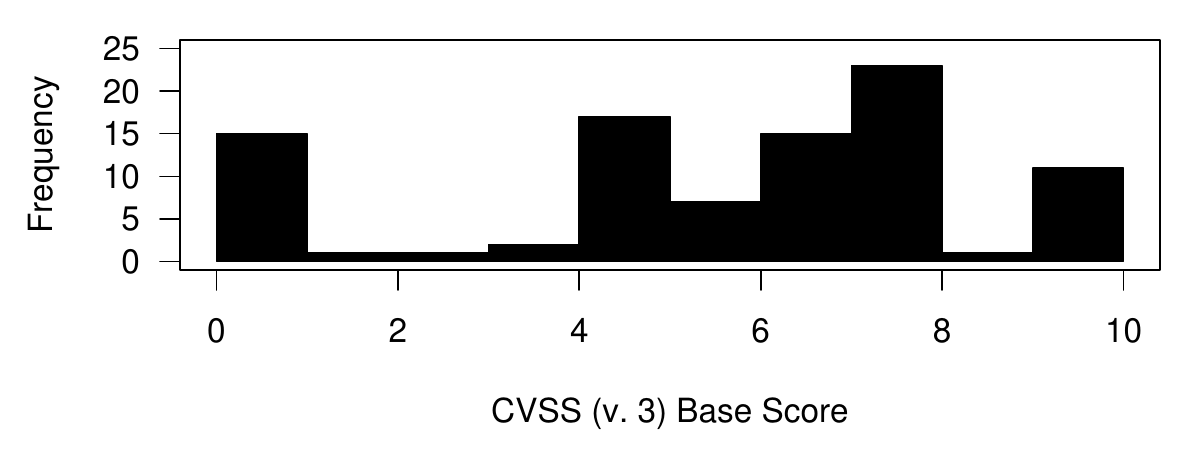}
\caption{Severity of the Vulnerabilities (CVSS)}
\label{fig: cvss}
\end{figure}

The empirical analysis can be started by taking a brief look at the volume of
vulnerabilities across time. Thus, Fig.~\ref{fig: years} displays the annual
vulnerability counts. As can be seen, the amounts of vulnerabilities reported
have steadily increased from the mid-2000s. The mean is about six
vulnerabilities per year. Many of the vulnerabilities have also been rather
severe, as can be concluded from Fig.~\ref{fig: cvss}. About 13\% of the
vulnerabilities have CVSS (v.~3) base scores higher than eight. The median is
close to six. The severity values are higher than what have been reported for
software written in Python~\cite{Ruohonen18IWESEP}, although not substantially
higher. The slight divergence is presumably explained by the C programming language.

\begin{figure}[th!b]
\centering
\includegraphics[width=\linewidth, height=4.5cm]{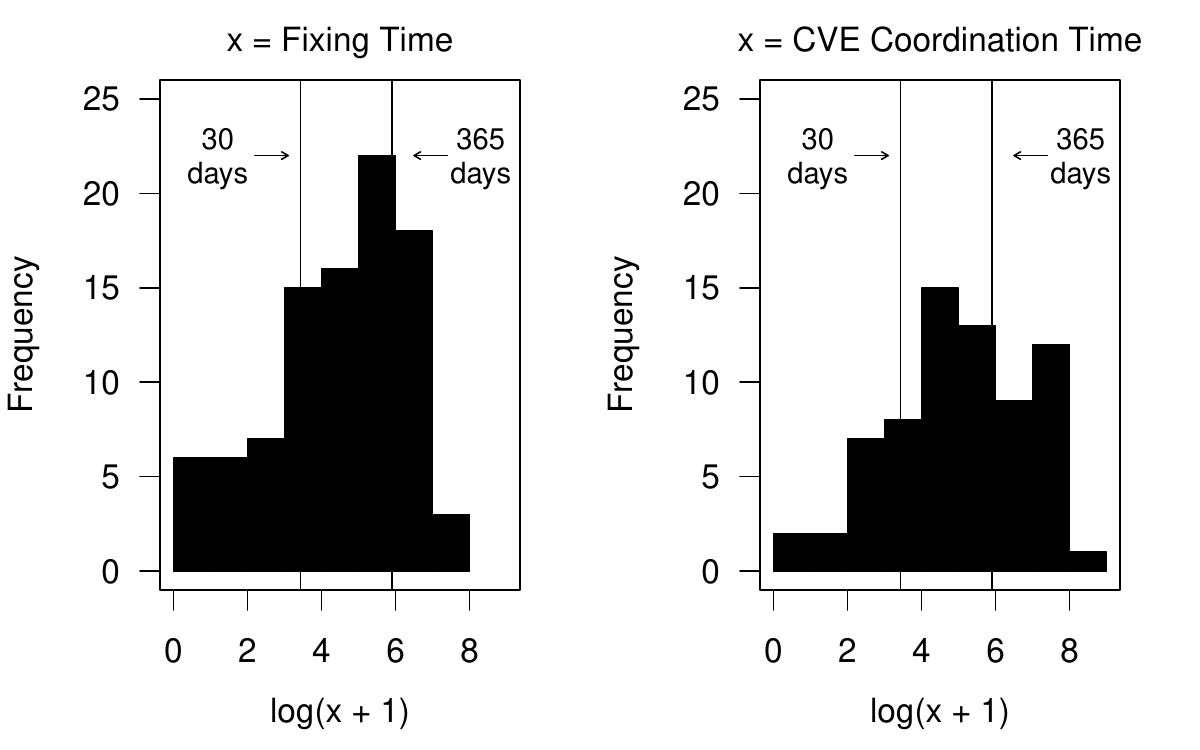}
\caption{Handling Times}
\label{fig: handling times}
\end{figure}

The two dependent variables are illustrated in Fig.~\ref{fig: handling times} by
using the noted logarithm transformation. As can be seen, both the vulnerability
fixing times and the CVE coordination times have been rather lengthy. In terms
of the former, the mean is $119$ days and the median is as long as $267$
days. The fixing times seem also longer than in software written in
Python. Although methodology is not directly comparable, previous studies have
reported that the median to fix vulnerabilities in packages distributed in PyPI
is about two months~\cite{Alfadel23}. Then, regarding the CVE coordination
times, the median is $157$ days. Although methodology is again different,
shorter timelines have been reported previously also in this
regard~\cite{Ruohonen18IST}. Having said that, nine vulnerabilities satisfied a
condition $(\textit{CVE Coordination Time})_i < 0$, meaning that CVEs were
already allocated before the vulnerabilities were disclosed to the PSF's
security team or the CPython's developers publicly.

\begin{table}[th!b]
\centering
\caption{$F$-tests for the OLS Regressions ($F$-values)}
\label{tab: reg}
\begin{tabular}{llccc}
\hline
&& Fixing time && CVE coordination time \\
\hline
REPORTER && 3.060$^{**\phantom{*}}$ && 0.917$^{\phantom{***}}$ \\
SEVERITY && 0.924$^{\phantom{***}}$ && 2.886$^{\phantom{***}}$ \\
POC && 0.386$^{\phantom{***}}$ && 0.241$^{\phantom{***}}$ \\
COMMITS && 0.204$^{\phantom{***}}$ && 0.011$^{\phantom{***}}$ \\
REFERENCES && 0.122$^{\phantom{***}}$ && 1.617$^{\phantom{***}}$ \\
COMMENTS && 1.560$^{\phantom{***}}$ && 0.064$^{\phantom{***}}$ \\
\hline
$n$ && $\phantom{0.0}$93$^{\phantom{***}}$ && 69 \\
R$^2$ && 0.923$^{\phantom{***}}$ && 0.842$^{\phantom{***}}$ \\
\hline
\multicolumn{5}{l}{~~~~~$^{***}~\textmd{for}~p < 0.001$, $^{**}~\textmd{for}~p < 0.01$, and $^{*}~\textmd{for}~p < 0.05$}
\end{tabular}
\end{table}

\begin{figure}[th!b]
\centering
\includegraphics[width=\linewidth, height=3cm]{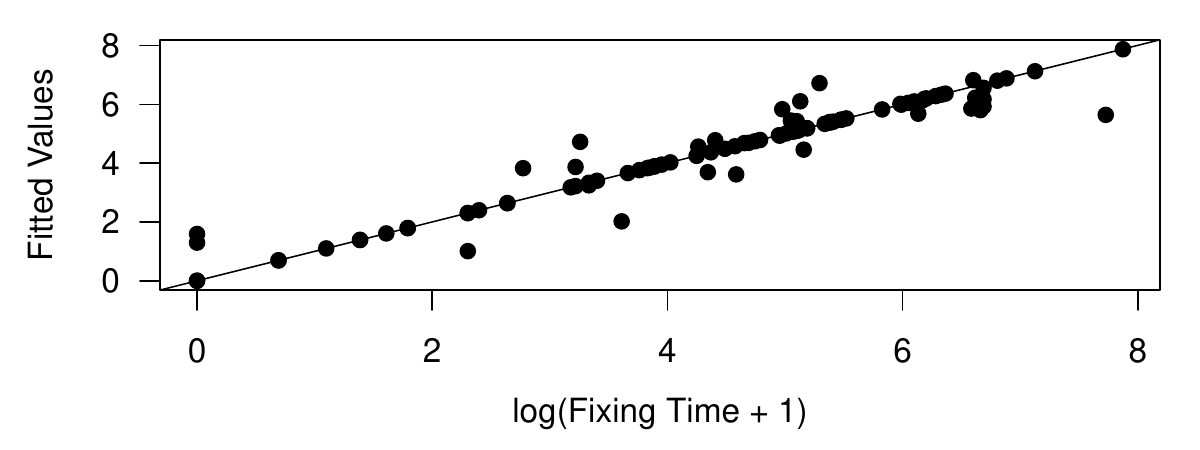}
\caption{Actual and Fitted Values from the OLS Regression for Fixing Times}
\label{fig: ols fit}
\end{figure}

Turning to the regression analysis, Table~\ref{tab: reg} summarizes the two OLS
regressions with $F$-tests. Unlike conventional \text{$t$-tests} for determining
the statistical significance of regression coefficients, a $F$-test determines
the statistical significance of whole variable groups, which is particularly
important regarding the dummy variables for the reporters of the
vulnerabilities. And, indeed, REPORTER is the only statistically significant
variable in the fixing time regression but not in the CVE coordination time
regression. Despite this general lack of statistical significance, model
performance is very good in both models. The fixing time regression yields a
coefficient of determination as high as $0.923$. The good performance is further
illustrated in Fig.~\ref{fig: ols fit}. However, the performance is almost
entirely because of the reporters. When fitting a new model with only REPORTER,
a value $R^2 = 0.909$ is obtained.

Basic graphical diagnostic plots indicate only modest concerns. On one hand,
normality of the residuals is an issue; Shapiro-Wilk tests~\cite{Shapiro65}
reject the null hypotheses that the residuals came from a normally distributed
population in both OLS models. Though, this commonplace normality violation is
debated, and the consequences may not be fatal at all~\cite{Knief21}. On the
other hand, heteroskedasticity is not a particular concern, unlike what has been
reported in previous studies with similar study
settings~\cite{Ruohonen20CHB}. According to two Breusch-Pagan
tests~\cite{BreuschPagan79}, the null hypotheses of homoskedasticity of the
residuals remain in force in both OLS models. Recomputing the models with the
Huber's \text{$M$-estimator} does not change the conclusion; only some reporter
dummy variables are statistically significant according to $t$-tests. Finally,
it should be mentioned that the models estimated are close to being ill-defined,
meaning that there are almost as many variables as there are observations. To
this end, a $R^2 = 0.256$ value is obtained for an OLS model including explicit
dummy variables for only those reporters who had reported at least two
vulnerabilities; all others are cascaded into a group of ``others''. Although
the value is much lower than in the previous models, it is still quite
reasonable for such a tiny regression model. Further including the remaining
variables does not improve performance; all of these other variables are
statistically insignificant, and $R^2$ actually decreases to $2.149$.

All in all, it can be concluded that some characteristics of people who report
CPython vulnerabilities largely explain the associated vulnerability fixing
times. A qualitative analysis would be likely needed to deduce about a
subsequent why-question. Based on the manual skimming of the associated bug
reports while constructing the dataset, it does not seem plausible that
communication and related things would explain the finding. Rather, it may be
that some people just write better vulnerability reports, perhaps providing
things that developers typically consider helpful, including reproducible tests,
sample code, and even patches~\cite{Bettenburg08, Karim19}. Alternatively, when
considering the bundling of third-party libraries, it may also be that the
faster fixing times have been due to people who are associated with Linux
distributions or other open source communities. In such a case, verification,
upstream coordination, and associated things may have already been done
elsewhere prior to handling a vulnerability at the CPython's
development~infrastructure.

\section{Discussion}\label{sec: discussion}

\subsection{Conclusion}

This short paper examined vulnerability handling times in CPython. Two variables
were specifically examined: vulnerability fixing times and CVE coordination
times. The paper's conclusions can be summarized with the following points:
\begin{itemize}
\itemsep 5pt
\item{CPython has seen a steady flow of new vulnerabilities, and the arrival
  rates have slightly increased from the 2000s. Many of the vulnerabilities are
  rather severe.}
\item{Both the vulnerability fixing times and the CVE coordination times have been lengthy. The medians are $267$ and $157$ days, respectively. In other words, it has taken on average about nine months to fix the vulnerabilities.}
\item{Based on regression analysis, only an identification of persons who have
  reported vulnerabilities is relevant statistically. In fact, merely including
  this identification data yields very good statistical performance. Contrary to
  many closely related previous studies, severity, proof-of-concept code,
  commits, comments posted to a bug tracker, and references to other sources
  explain neither the vulnerability fixing times nor the CVE coordination~times.}
\end{itemize}

In addition to these brief conclusive three points, some limitations should be
acknowledged. After elaborating these, a couple of points follow about the
potential for further research.

\subsection{Limitations}

Regarding limitations, the obvious needs to be explicitly mentioned: as only a
single case was analyzed, nothing can be deduced about vulnerability handling
times in other projects. Rather than trying to seek generalizability, which
already in the open source context is difficult, if not even impossible, it
might be more reasonable to address the limitation by examining other
interpreters for interpreted programming languages. Such an examination might,
or should, also reveal whether vulnerabilities are similar or different across
interpreters.

Also construct validity and robustness of the data might be slightly threatened.
As was described in detail in Section~\ref{sec: data and methods}, the presence
of two vulnerability tracking systems complicated the data collection
process. In addition, many concessions had to be made when operationalizing the
variables for the analysis. However, these problems are fairly typical to the
research domain. Already the fundamental abstract question of what counts as a
vulnerability is not straightforward; a single CVE may reference multiple
vulnerabilities and the other way around, at least according to the CPython's
developers. It should be also acknowledged that manual data collection often
involves a degree of subjectivity; in this regard, particularly the POC
independent variable may be subjected to criticism.

\subsection{Further Work}

The paper's main result---the importance of reporters and their
characteristics---would deserve a closer look in further research. Why the
vulnerability fixing times can be predicted so well by merely identifying the
reporters of the CPython vulnerabilities? Although a similar observation has
been made also in previous studies, the answers to the question have been only
tentative and without much theoretical contributions. If expertise of reporters,
including with respect to providing reproducible information, is as important as
often seen in the literature~\cite{Karim19}, it might be also possible to make
practical advances by providing better instructions to people about good
vulnerability reporting practices. At the moment, the PSF's policy~\cite{PSF24a}
does not say anything about reproducibility, POCs, and other similar things
related to vulnerability handling.

Another good question for further research would be to examine a third delay metric:
the time required to integrate the fixes to releases. According to the CPython's
old vulnerability tracker, it seems that such integration too has taken a rather
long time. The many software engineering work activities related to
integration~\cite{Adams16b} may offer an explanation. However, it may well also
be that new CPython releases are not pushed forward merely to address
vulnerabilities. That is, the vulnerability fixes may be postponed to match
existing release engineering schedules and plans. Whatever the explanation may
be, the integration timelines are relevant in terms of actual security risks
because users typically only get vulnerability fixes through releases, meaning
that they are often running vulnerable Python interpreters until a new release
is made. In fact, one could further extend the examination toward observing
delays that occur until third-party distributors have integrated the CPython
releases into their package managers.

\balance
\bibliographystyle{abbrv}

\end{document}